\documentclass[12pt,twoside,a4paper]{article}
\usepackage{amsmath,amssymb,latexsym,theorem,natbib,epsfig,color,subfigure,footmisc,bbm}
\usepackage{multirow,graphicx,array,rotating,url,multicol}  
\usepackage{epstopdf,afterpage,wasysym,hyperref}
\usepackage{verbatim}
\usepackage[normalem]{ulem}
\def\crps{\mathop{\hbox{\rm CRPS}}}

\def\ri{\mathop{\hbox{\rm RI}}}
\def\qs{{\mathrm {QS}}}

\title{AI and physics-based weather forecasting: A comparative study}

\author{{M\'aty\'as Kocsis}$^{1,2}$ and {S\'andor Baran}$^{1}$ \vspace*{0.5cm}\\
{\small $^1$Faculty of Informatics, University of Debrecen, Hungary} \\
{\small $^2$Doctoral School of Informatics, University of Debrecen, Hungary} 
}

\date{}

\begin{document}
\maketitle

\begin{abstract}
In the last few years, artificial intelligence-based models have become the centre of attention in weather forecasting due to their increasing accuracy and efficiency. Pioneering among weather services, the European Centre for Medium-Range Weather Forecasts (ECMWF) has developed its Artificial Intelligence Forecasting System (AIFS) model, which was first to provide data-driven ensemble forecasts in June 2024. Since July 2025, the AIFS ensemble model has been operational and runs in parallel with ECMWF’s physics-based Integrated Forecasting System (IFS), which is considered the gold standard in weather prediction. The new AIFS model can generate forecasts ten times faster than the classical numerical weather prediction model, while consuming approximately a thousand times less energy.
We present the results of our systematic assessment of the performance of the IFS and AIFS models by comparing the accuracy of raw and post-processed medium-range 10-m wind-speed ensemble forecasts generated operationally by the two models for the period between July and November 2025 for more than 9000 synoptic observation stations across the globe. The post-processed case involves the parametric ensemble model output statistics (EMOS) as well as the non-parametric quantile regression (QR) approach to correct any systematic biases and dispersion inaccuracies in the raw forecasts.
The predictive performance of raw IFS ensemble forecasts proves to be substantially superior to the skill of the raw AIFS predictions for all investigated forecast horizons. As expected, post-processing significantly improves the skill of both IFS and AIFS predictions, and, across most verification metrics, EMOS is superior to QR, especially for short lead times. 
Compared to the raw ensemble, the differences in skill between the matching IFS and AIFS predictions are substantially decreased by post-processing and are mostly significant at short lead times, when the IFS forecasts outperform their AIFS counterparts.

\bigskip
\noindent {\em Keywords:\/} artificial intelligence-based forecasting, ensemble model output statistics,  ensemble post-processing, numerical weather prediction, quantile regression
\end{abstract}
 
\section{Introduction}
\label{sec1}

Since their first operational implementations in the 1950s, numerical weather prediction (NWP) models have become the primary tools of weather forecasting \citep{NWP50}. These mathematical models describe the dynamical and physical behaviour of the atmosphere and the coupled systems, and forecasts are obtained by solving the corresponding sets of non-linear partial differential equations numerically. The next crucial shift in the practice of weather prediction was the introduction of ensemble prediction systems (EPSs) in operational use in 1992 at the European Centre for Medium-Range Weather Forecasts (ECMWF) and the National Centers for Environmental Prediction (NCEP), followed shortly by the Meteorological Service of Canada \citep{b18a}.
By running an NWP model several times with different initial conditions and/or model parametrizations, a forecast ensemble captures forecast uncertainty and estimates predictive distribution of the weather variable at hand, thereby enabling probabilistic weather forecasting \citep{gr05}. In the last three decades, ensemble forecasting has become a routine all over the world, and all major weather services have their own operational EPS. One of the most prominent examples is the ECMWF's Integrated Forecasting System \citep[IFS;][]{ecmwf24}, which since the introduction of Cycle 48r1 on 27 June 2023, generates global 51-member medium-range forecasts up to 15 days ahead and 101-member sub-seasonal-range predictions covering forecast horizons of 1--46 days both at 137 vertical model levels and at horizontal resolutions of approximately 9 km ($\text{T}_{\text{CO}}1279$) and 36 km ($\text{T}_{\text{CO}}319$), respectively. Medium-range forecasts are produced four times a day with 00, 06, 12, and 18 UTC initializations, whereas sub-seasonal-range predictions run from 00 UTC, both requiring a vast amount of computational power.

Recently, physics-based NWP models have received serious competitors in machine learn\-ing (ML)-based weather models, developed mostly by large tech companies, like the Pangu-Weather by Huawei \citep{bi23}, the GraphCast by Google DeepMind \citep{lam23}, or the FourCastNet by NVIDIA \citep{kurth23}. The performance of these data-driven approaches producing deterministic forecasts, all trained on the ECMWF's ERA5 reanalysis archive \citep{era5}, are comparable with the deterministic (control) forecasts of the ECMWF IFS at a negligible computational cost. In response to challenges, ECMWF also started to develop the Artificial Intelligence/Integrated Forecasting System \citep[AIFS;][]{aifs24} and became a pioneer in operational use of ML-based weather forecasts. The deterministic model (AIFS Single) has been running operationally since 25 February 2025, whereas the operational version of the 51-member  probabilistic model \citep[AIFS-CRPS;][]{aifs-crps26}, replacing the diffusion-based experimental approach (AIFS-DIFF), was released on 1 July 2025. Both AIFS models run daily at  00, 06, 12, and 18 UTC with a forecast range up to 15 days at a vertical resolution of 13 model levels and at horizontal resolutions of approximately 30 km. Note that several months after the experiments with the AIFS-DIFF model had started, in December 2024, Google DeepMind also introduced its data-driven ensemble model GenCast \citep{gencast25}, with a forecast skill comparable with the ECMWF IFS.

Despite the continuous efforts of weather centers to improve their EPSs, ensemble forecasts may still be subject to systematic errors, such as bias or lack of calibration \citep[see e.g.][]{bhpt05, ecmwfeval25}. A possible direction towards correcting such problems is the use of some form of post-processing \citep{b18b}. Over the past few decades, a large palette of post-processing methods has been developed for a wide range of weather quantities, comprising both statistical and ML-based models; for comprehensive overviews see e.g. \citet{w18} or \citet{vbd21}. These approaches can either be parametric, resulting in full predictive distributions, or non-parametric, estimating the quantiles of the predictive distribution or providing adjusted ensemble forecast. Prominent representative of the parametric methods are
the ensemble model output statistics \citep[EMOS;][]{grwg05}, the Bayesian model averaging \citep[BMA;][]{rgbp05} or the artificial neural network (ANN)-based distributional regression network \citep[DRN;][]{rl18}, while non-parametric techniques cover variations of quantile regression including statistical approaches \citep[see e.g.][]{fh07,brem19} and ML-based models \citep[see e.g.][]{b20bqn,sy24ncqrnn}, quantile mapping \citep{hs18}, member-by-member post-processing \citep{vsv15}, random forests \citep{tmzn16qrf}, or ANN-based forecast adjustment \citep{bmcd25}. In general, non-parametric models are often superior to parametric ones \citep[see e.g.][]{mblhyb26}, furthermore, due to their flexibility in accommodating various weather quantities as input features and capability of modelling complex non-linear relationships, ML models usually outperform traditional statistical methods; for recent systematic comparisons we refer to \citet{sl22} and \citet{pslh26}.

The present work aims to provide a systematic comparison of the predictive performance of raw and post-processed operational physics-based (IFS) and data-driven (AIFS-CRPS) ECMWF 10-m wind-speed forecasts using data for the period spanning from 1 July (operational launch of the AIFS-CRPS) to 30 November 2025. For ensemble calibration, we utilize two different approaches. On the one hand, we consider the parametric EMOS approach, based on a truncated normal (TN) predictive distribution \citep{tg10}. On the other hand, we apply a non-crossing version of the flexible non-parametric quantile regression \citep{mstc13}, which does not require preliminary assumptions about the family of the predictive law. To our best knowledge, this is the first study where the skill of post-processed AIFS-CRPS ensemble forecasts is investigated; however, a similar comparison of raw medium-range IFS and AIFS-CRPS 10-m wind speed ensemble predictions for a different time period (1 February to 30 September 2024) can be found in \citet{aifs-crps26}.

The paper is organized as follows. Section \ref{sec2} is devoted to the detailed description of the wind-speed data used in this study. Section \ref{sec3} provides details of the considered post-processing methods, together with the approaches to training data selection and tools applied for forecast evaluation. Section \ref{sec4} reports the main results, followed by Section \ref{sec5}, which discusses our findings and contains our concluding remarks. Additional results on spatial patterns in relative performance of matching IFS and AIFS forecasts can be found in Appendix \ref{secA}.

\section{Data}
\label{sec2}

\begin{figure}[t!]
    \centering
    \includegraphics[width=.85\textwidth]{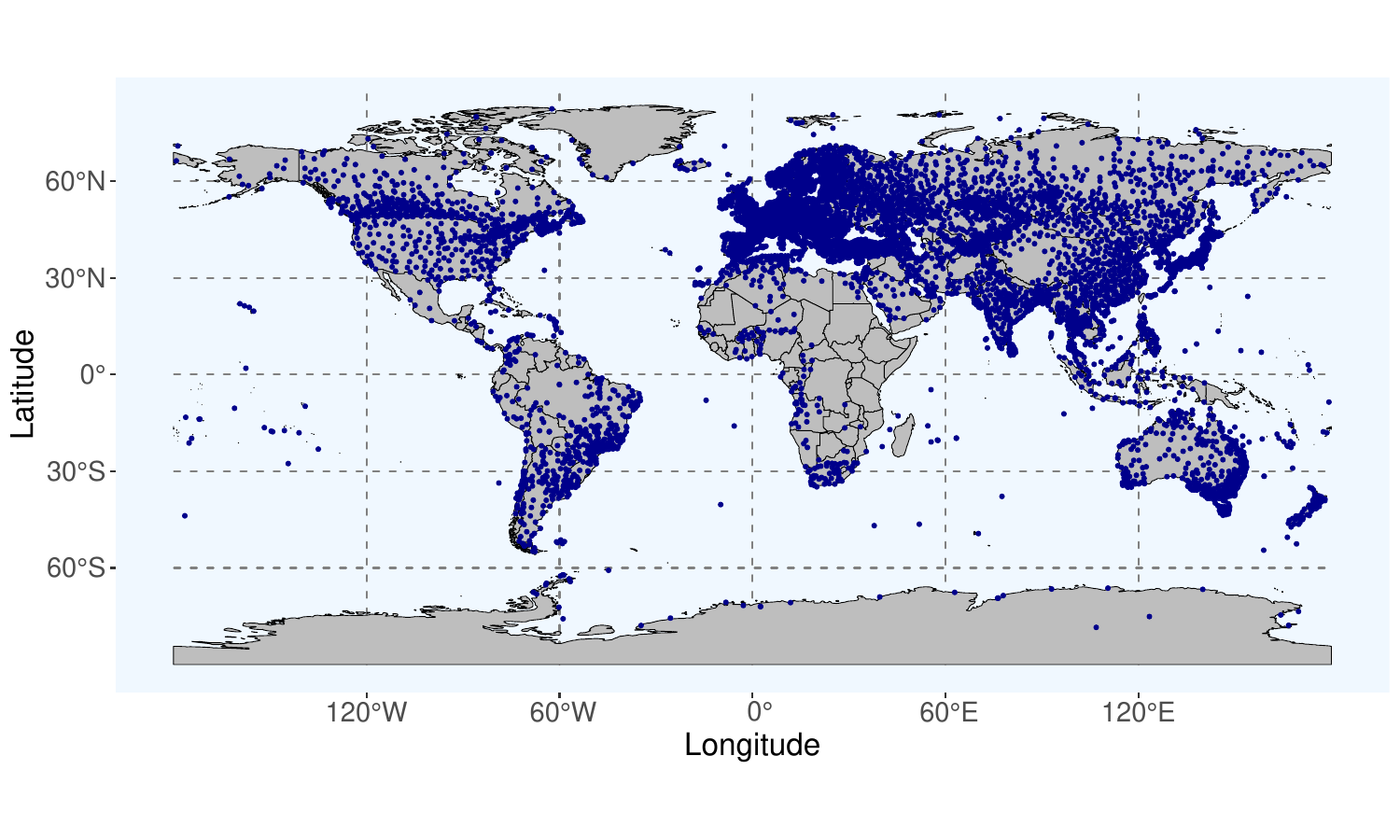}
    \caption{Location of SYNOP stations}
    \label{fig:map}
\end{figure}

We consider 50-member 24, 48, \ldots , 360h ahead operational ensemble forecasts of 10-m wind-speed produced by the ECMWF IFS (cycle CY49R1) and AIFS-CRPS (AIFS ENS v1) systems together with the corresponding validating observations for 9246 SYNOP stations (see Figure \ref{fig:map}). Note that both for IFS and AIFS, forecasts at SYNOP stations are predictions for the nearest grid point, while observations at the reported observation time are 10-minute averages of the observed 10-m wind speed. The 50 ensemble members in both cases are generated using perturbations in the initial conditions; thus, they are statistically indistinquishable. All forecasts are initialized at 12 UTC and cover a period of five months, from 1 July to 30 November 2025 (153 calendar days). Forecast-observation pairs were subject to  quality control, and only those locations were kept from a total set of 16367 sites where the proportion of dates with missing data does not exceed 5\,\%. 

\section{Post-processing methods and forecast evaluation tools}
\label{sec3}

In what follows, let \ $f_1, f_2, \ldots , f_{50}$ \ denote the 50-member (IFS or AIFS) 10-m wind-speed ensemble forecast for a given location with a given initialization date and forecast horizon, and denote by \ $\overline f$ \ and \ $S^2$ \ the ensemble mean and variance, respectively, that is
\begin{equation*}
  \overline f := \frac 1{50} \sum _{k=1}^{50} f_k \qquad \text{and} \qquad S^2:= \frac 1{49}\sum_{k=1}^{50}\big(f_k - \overline f\big)^2.
\end{equation*}

As mentioned in the Introduction, for post-processing wind-speed ensemble forecasts, we consider two conceptually different approaches: the parametric TN EMOS detailed in Section \ref{subs3.1}, and non-crossing quantile regression introduced in Section \ref{subs3.2}.

\subsection{Ensemble model output statistics}
\label{subs3.1}

Ensemble model output statistics (or non-homogeneous regression) is a simple, but powerful parametric post-processing method, often serving as a benchmark calibration approach. The EMOS predictive distribution is a single parametric law with distributional parameters expressed as simple functions of the corresponding ensemble forecasts. Thus, EMOS models for various quantities differ in the considered parametric family, which should address the properties of the predictand, and the link functions between the ensemble members and the distributional parameters. For instance, temperature is mainly modelled by a Gaussian distribution \citep{grwg05}, whereas wind speed can be described by a right-skewed law with positive support. \citet{jhmd78} proposed a Weibull distribution, \citet{gtpf98} found a gamma law to be appropriate, which distribution was also later used in BMA modelling of wind speed \citep{sgr10}. In the context of EMOS modelling, truncated normal, log-normal \citep{bl15}, generalized extreme value \citep[GEV;][]{lt13}, and truncated GEV  distributions \citep{bszsz21} have been investigated so far.

As mentioned, here we follow the EMOS approach of \citet{tg10}, where the predictive distribution is a Gaussian law left-truncated at zero \ ${\mathcal N}_0\big(\mu,\sigma ^2\big)$ \ with location \ $\mu$ \ and scale \ $\sigma >0$. \ The location and scale parameters are linked to the forecast ensemble as
\begin{equation*}
  \mu = a + b^2\overline f \qquad \text{and} \qquad \sigma^2=c^2 + d^2 S^2,
\end{equation*}
giving equal weights to the exchangeable ensemble members, as 
suggested by \citet{gneiting14}. In line with the optimum score estimation principle of \citet{gr07}, model parameters \ $a, \ b, \ c, \ d \in{\mathbb R}$ \ are estimated by optimizing a proper scoring rule over training data consisting of past forecasts of 10-m wind-speed and corresponding validating observations. The most popular choice is the continuous ranked probability score (CRPS) defined by \eqref{eq:crps} and discussed in detail in Section \ref{subs3.4}.

\subsection{Quantile regression}
\label{subs3.2}

Since its introduction in 1978, quantile regression (QR) has become an indispensable tool in modern regression analysis \citep{k05qr}. The primary distinction between this method and least squares estimation is that, based on explanatory variables, quantile regression models the conditional quantiles of the response variable. This is achieved by minimizing the sum of {\em pinball losses\/}, defined by \eqref{eq:pinball}, over the training data, which can be framed as a linear programming problem and solved by efficient algorithms.

Due to its robustness, QR has been successfully applied in many areas, including ensemble post-processing. The main advantage of this method over parametric techniques is that no distributional assumptions are required on the weather or other quantity of interest. Many times in practice, the distribution of the predicted quantity is not well described by any parametric distribution; thus, QR has the potential to provide better calibrated forecasts than parametric methods \citep[for a recent comparison in the field of photovoltaic energy forecasting, see][]{mblhyb26}.

Here, we consider an approach similar to that of \citet{brem19}. Let \ $q_{10}, \ q_{50}$, \ and \ $q_{90}$ \ denote the $10$th, $50$th, and $90$th percentiles of the raw ensemble forecast, respectively, and let \ $y\geq 0$ \ denote the 10-m wind speed. To ensure nonnegative predictions, the natural logarithm of the 10-m wind speed is modelled. More specifically, for each \ $\tau\in\big\{1/51, 2/51,\ldots, 50/51\big\}$, \ the conditional \ $\tau$ \ quantile of \ $z:=\log(y + 0.01)$ \ with respect to the forecast ensemble is modelled as
\begin{equation*}
  q_{\tau}(z \mid f_1, f_2,\ldots, f_{50}) = \beta_{0,\tau} + s_{1,\tau}(q_{10}) + s_{2,\tau}(q_{50}) + s_{3,\tau}(q_{90}),
\end{equation*}
where \ $\beta_{0,\tau}\in\mathbb{R}$, \ and for each \ $k\in\{1,2,3\}$ \ the function \ $s_{k,\tau}$ \ is a cubic spline having the form
\begin{equation*}
  s_{k,\tau}(x) = \sum_{j = 1}^{J} b_{k,\tau,j} B_{k,j}(x),
\end{equation*}
	with \ $J\in\mathbb{N}, \ b_{k,\tau} = (b_{k,\tau,1},\dots,b_{k,\tau,J})\in\mathbb{R}^{J}$, \ and \ $\{B_{k,j}\}_{j = 1}^{J}$ \ is the corresponding spline basis.

In the objective function to be minimized to estimate model parameters, besides the pinball loss, a penalty term on the coefficients is also introduced to control the shape of the splines, and hence, prevent the model from overfitting. More specifically, one has to minimize
\begin{equation}
\label{eq:qrloss}
\mathcal{L}(\beta_{0,\tau}, b_{1,\tau}, b_{2,\tau}, b_{3,\tau}) := \sum_{i=1}^{N} \rho_{\tau}\big(z_i - q_{\tau}(z_i \mid f_{i,1}, f_{i,2},\dots, f_{i, 50})\big)
		+ \lambda\sum_{k=1}^3 \sum_{j=1}^{J - 3} |\Delta^3 b_{k,\tau,j}|,
              \end{equation}
where \ $N$ \ is the total number of forecast cases in the training data, \ $z_i$ \ denotes the $i$th log-transformed observation, \ $f_{i,1},f_{i,2}, \ldots,f_{i,50}$ \ is the corresponding ensemble forecast,  \ $\rho_\tau(x)$ \ is the pinball loss function
\begin{equation}
    \label{eq:pinball}
\rho_\tau(u) := 
\begin{cases}
	u \tau,  & \text{if  \ $u \geq 0$,} \\
		u (\tau-1), & \text{if  \ $u < 0$,}
        \end{cases}
      \end{equation}
      $\lambda > 0$ \ is the penalty coefficient, and \ $\Delta^3$ \ denotes the third-order difference operator.

A common issue when estimating multiple quantiles is quantile crossing, a phenomenon that limits the interpretability of the results. Since the models are fit independently for each quantile level, the obtained quantile curves might intersect, contradicting the fact that the predictive cumulative distribution function (CDF) is monotone increasing. In our approach, non-crossing quantile estimations are guaranteed by suitable additional constraints introduced to the linear programming task performing the minimization of the objective function \eqref{eq:qrloss}, see \citet{mstc13}.      

\subsection{Model training}
\label{subs3.3}

The predictive performance of any post-processing model highly depends on the temporal and spatial decomposition of the training data. In the case of ML-based techniques, usually a very long, but static training window is considered \citep[see e.g.][]{rl18,sl22}, whereas in EMOS modelling, it is common to use sliding training windows \citep{grwg05, tg10} based on forecast-observation pairs for the preceding \ $n$ \ calendar days. This latter method allows a quick adaptation, for instance, to seasonal changes; however, the sliding window should be wide enough to allow a stable parameter estimation. For a detailed comparison of time-adaptive training data selection methods, we refer to \citet{llmssz20}.

In terms of spatial composition, a popular choice is local modelling, when the post-processing model for a given location is estimated using solely forecasts  and observations for that particular place. On the opposite side of the palette stands the global (or regional) approach, which relies on data for all considered locations \citep{tg10} and results in a single model for the entire ensemble domain. In general, local models are superior to their regional counterparts, provided the training period is long enough to avoid numerical problems in the modelling-related optimization process. The optimal training period length for EMOS models for temperature, wind speed, and precipitation accumulation can be found in \citet{hemri14}. Regional modelling requires far shorter training windows; however, it is usually not suitable for heterogeneous domains covering large territories or the whole globe. Finally, as a reasonable trade-off between local and regional modelling, one can consider the clustering-based semi-local approach of \citet{lb17}. First observation stations are clustered based on features representing the station climatology and/or the forecast error of the ensemble mean during the training period, and then regional modelling is performed within each cluster. This training technique proved to be successful for a large variety of weather quantities and ensemble domains, see, for instance, \citet{szgb23} or \citet{bl24}.

\subsection{Forecast evaluation}
\label{subs3.4}

According to the suggestions of \citet{gbr07}, forecast evaluation should be consistent with the main goal of probabilistic forecasting, which is ``maximising the sharpness of the predictive distributions subject to calibration''. Sharpness refers to the spread or concentration of probabilistic forecasts and is a property of the forecast itself. Calibration, on the other hand, refers to the statistical consistency between the predicted and observed probabilities of events. Thus, calibration is a joint property of the forecast-observation pairs. Forecast skill is usually assessed by various graphical tools and with the help of scoring rules, which are functions assigning positive values to (probabilistic or point) predictions and corresponding validating observations.

An easily interpretable graphical tool for investigating calibration of ensemble forecasts or samples drawn from a predictive distribution is the verification rank histogram or Talagrand diagram \citep[see][Section 9.7.1]{w19}. It is the histogram of ranks of the observations with respect to the corresponding forecasts, with ties resolved randomly. In the case of a properly calibrated $K$-member forecast, these ranks should be uniformly distributed on the set \ $\{1, 2, \ldots ,K+1\}$ \ of possible ranks, resulting in flat histograms. Biased forecasts lead to triangular shapes, while over- and underdispersion produce hump- and U-shaped rank histograms, respectively. Besides visual perception, one can also quantify the deviation from uniformity with the reliability index \citep[RI;][]{dmhzds06}, which allows a ranking of the competing forecasts. The RI is defined as 
\begin{equation*}
   \label{eq:relind}
 \ri:=\sum_{r=1}^{K+1}\Big| \rho_r-\frac 1{K+1}\Big|,
\end{equation*}
where \ $\rho_r$ \ denotes the relative frequency of rank \ $r$ \ over all considered forecast cases in the verification period.

To assess calibration, one can also consider the coverage of \ $(1-\alpha)\times 100\,\%, \ \alpha \in (0,1),$ \ central prediction intervals, which is the percentage of observations lying between the lower and upper \ $\alpha/2$ \ quantiles of the predictive distribution. For a calibrated forecast, the coverage should be around \ $(1-\alpha)\times 100\,\%$, \ and level \ $\alpha$ \ is usually chosen to match the nominal coverage of the raw ensemble, which for a \ $K$-member probabilistic forecast equals \ $(K-1)/(K+1) \times 100\,\%$. \ Furthermore, the average width of these central prediction intervals can serve as a measure of sharpness.

Calibration and sharpness can be assessed simultaneously using the continuous ranked probability score \citep[CRPS;][Section 9.5.1]{w19}, which quantifies the deviation between the predictive CDF 
and the empirical CDF of the corresponding observation. In particular, for an observation \ $y \in {\mathbb R}$ \ and predictive CDF \ $F$, 
\begin{equation}
  \label{eq:crps}
  \crps(F, y) := \int_{-\infty}^{\infty} \left( F(z) - \mathbbm{1}\{z \geq y\} \right)^2 \, {\mathrm d}z, 
\end{equation}
where \ $\mathbbm{1}\{\cdot \}$ \ denotes the indicator function. The CRPS is a negative-oriented score, that is, the smaller the better; moreover, for the left-truncated Gaussian distribution, it has a closed form \citep{tg10}, which allows an efficient optimization process in TN EMOS modelling. Finally, for a forecast ensemble, or when one deals with a sample drawn from the predictive distribution, the predictive CDF \ $F$ \ in \eqref{eq:crps} should be replaced by the corresponding empirical one, see e.g. \citet{jkl19}.

To evaluate the forecast skill of quantile regression, resulting in a representation of a predictive CDF \ $F$ \ in the form of its \ $\tau$-quantiles
\begin{equation*}
  q_\tau(F) := F^{-1}(\tau):= \inf\{y:F(y)\geq\tau\}, \qquad 0 < \tau < 1,
\end{equation*}
a popular negatively oriented scoring rule is the quantile score \citep[QS; see e.g.][Section 9.6.1]{w19}. Given an observation \ $y$, \ the QS is defined as
\begin{equation}
    \label{eq:QSdef}
\qs _{\tau}(F,y):=\rho_\tau\big(y -q_{\tau} (F) \big),
\end{equation}
where \ $\rho_\tau(x)$ \ is the pinball loss function defined by \eqref{eq:pinball}. Similar to the CRPS, QS is a proper scoring rule, and in Section \ref{sec4}, summarizing our findings, we consider QS scores for the 5th, 10th, 20th, 80th, 90th, and 95th  quantiles of the competing probabilistic forecasts.

Finally, as a point forecast we consider the median of the predictive distribution, which is evaluated using the mean absolute error (MAE).

Besides considering the mean value \ $\overline{\mathcal S}_F$ \ of a score \ $\mathcal S$ \ for a forecast \ $F$ \ over the verification data, one is often interested in the improvement of \ $F$ \ in terms of the given score with respect to a reference forecast \ $F_{\text{ref}}$, \ resulting in mean score values  \ $\overline{\mathcal S}_{F_{\text{ref}}}$, \ as it might provide a more detailed comparison of the various predictions. This improvement can be quantified with the help of a skill score \citep{murphy73}, defined as
\begin{equation*}
  \label{eq:skillscore}
  \mathcal{SS}\big(F;F_{\text{ref}}\big) := 1 - \frac{\overline{\mathcal S}_F}{\overline{\mathcal S}_{F_{\text{ref}}}},
\end{equation*}
where larger values indicate better skill. In Section \ref{sec4} we report the continuous ranked probability skill score (CRPSS) and quantile skill score (QSS) for probabilistic, and MAE-based skill score (MAES) for point forecasts.

Significance of differences in score values corresponding to different forecasts is addressed by enhancing the skill score values with 95\,\% Gaussian confidence intervals calculated using 2000 block bootstrap samples. First, we sample the spatially averaged score value of the actual and reference forecasts using the stationary block bootstrap scheme of \citet{pr94}, where the block length follows a geometric distribution. Then we calculate the 2000 skill score values from the means of the matching samples.

\subsection{Implementation details}
\label{subs3.5}

Both EMOS and QR post-processing models are fit separately for all lead times using 60-day sliding training windows. This training period length, in general, provides enough training data for stable parameter estimation while keeping an appropriate number of test forecast cases for reliable assessment of model performance. Note that shorter training periods were also tested; however, for instance, a 30-day training window resulted in unstable models at lead times beyond 10 days.

In the case of EMOS models, local training is applied by minimizing the cumulated CRPS over the training period. However, in some cases, usually when extremely high wind-speed observations are present in the training data, the estimations apt to result in negative and in absolute value large location parameter along with a large scale, which completely distorts the model performance. This issue is handled by an L$2$ regularization of the parameters. Namely, if the sum of the absolute values of the estimated parameters exceeds $50$, the loss function is extended with an additive regularization term, which equals 0.1\,\% of the sum of squares of the parameters.

QR models are fit using a clustering-based semi-local approach, as these models possess significantly more parameters compared to EMOS. At each forecast date, stations are classified into $100$ clusters using the k-means algorithm based on the observations and forecast errors of the preceding 60 days. More specifically, the cluster analysis is based on $24$ features assigned to each station, the first half of which contains equidistant quantiles of the climatological CDF over the 60-day training period, while the remaining components consist of equidistant quantiles of the empirical CDF of forecast error of the ensemble mean. We have tested several settings for the number of clusters, from $20$ up to $500$, and $100$ was found to be an appropriate choice regarding model performance. To fit the quantile regression model described in Section \ref{subs3.2}, the {\tt quantregGrowth} package \citep{mtesa21}of {\tt R} was used, which has been specifically developed for estimating penalized quantile regression splines. To decrease computation costs, the smoothing parameter \ $\lambda$ \ is fixed to one, while the number of nodes for the splines is chosen according to the default settings depending on the size of the training sample. To handle unsuccessful parameter estimation, we applied a fall-back model with \ $\lambda =1.2$ \ and with a fixed number of five nodes, suggested by the authors of the packages. Nevertheless, this fall-back model was used (and occurred to be successful) only in around $0.03\,\%$ and $0.02\,\%$ of the forecast cases for the IFS and AIFS, respectively.

\section{Results}
\label{sec4}

\begin{figure}[t]
\begin{center}
\epsfig{file=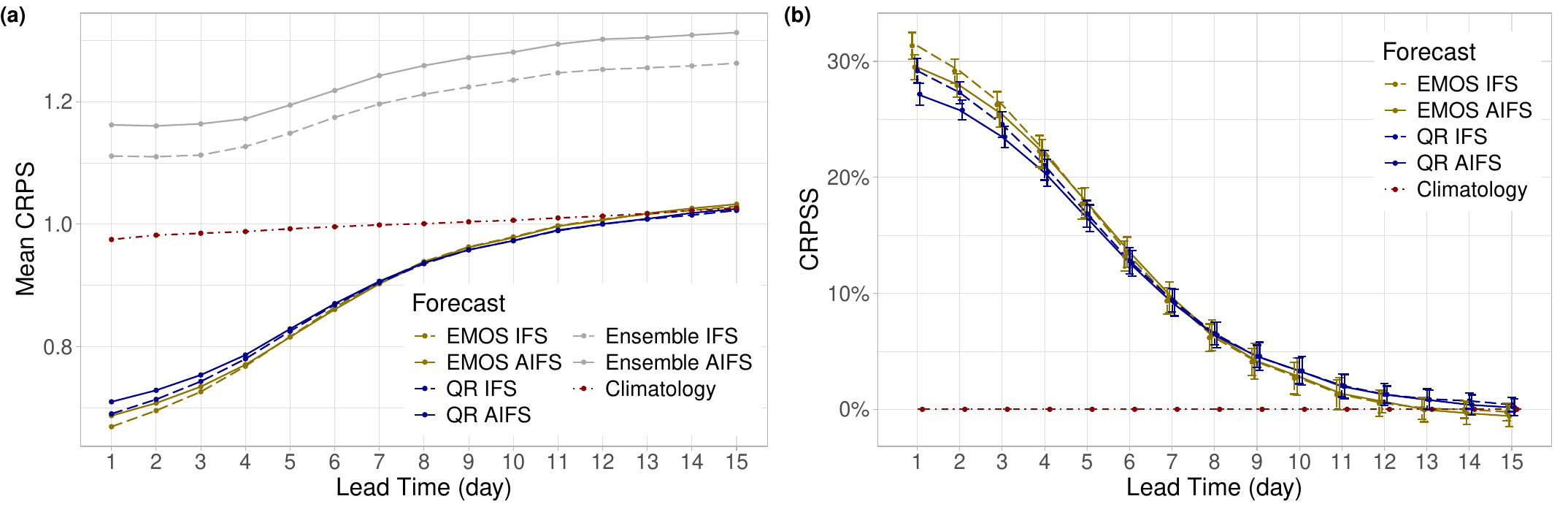, width=\textwidth}
\end{center}
\caption{Mean CRPS of post-processed, raw and climatological wind-speed forecasts (a) and CRPSS of post-processed forecasts with respect to climatology together with 95\,\% confidence bounds (b) as functions of the lead time.}
\label{fig:crps_crpss}
\end{figure}

In what follows, we provide a detailed evaluation of the forecast skill of raw and post-processed IFS and AIFS predictions using forecast-observation pairs for the period from 13 September to 30 November 2025 (79 calendar days, immediately following the first 60-day training window used in EMOS and QR modelling). To ensure fair comparability of the QR method with the parametric EMOS approach, resulting in full predictive laws, for the latter, we consider 50 equidistant quantiles of the corresponding TN predictive distributions. Furthermore, climatological forecasts used as baseline predictions are also based on the most recent 50 observations of wind-speed.

\begin{figure}[t]
\begin{center}
\epsfig{file=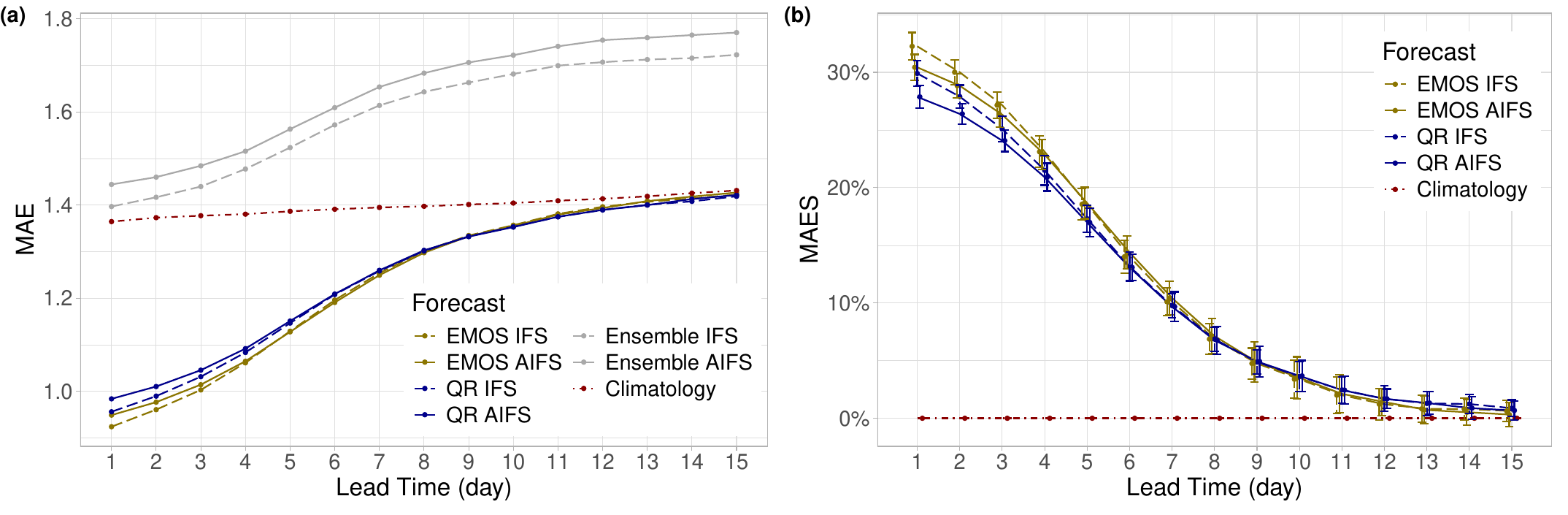, width=\textwidth}
\end{center}
\caption{MAE of post-processed, raw and climatological median forecasts (a) and MAES of post-processed forecasts with respect to climatology together with 95\,\% confidence bounds (b) as functions of the lead time.}
\label{fig:mae_maes}
\end{figure}

According to Figure \ref{fig:crps_crpss}a, in terms of the mean CRPS, raw IFS forecasts considerably outperform their AIFS counterparts, and the difference is nearly identical along all forecast horizons. Both raw ensemble predictions are behind climatology; however, this is in line with the findings of \citet{bszsz21} based on IFS wind-speed forecasts for more than 1000 SYNOP stations in Europe and Asia covering more than 1500 calendar days. As expected, both post-processing methods for both EPSs substantially decrease the mean CRPS, and as Figure \ref{fig:crps_crpss}b indicates, up to day 11, all calibrated predictions are significantly superior to climatology. For short lead times, IFS maintains its advantage over AIFS after calibration, too; however, the difference between the corresponding IFS and AIFS predictions is minor and gradually decreases with the forecast horizon. From the two investigated calibration techniques, the parametric EMOS is consistently ahead of the non-parametric QR till day 7, both for IFS and AIFS, but the difference is significant only at the first few lead times (days 1 -- 3 for IFS; days 1 -- 4 for AIFS).

\begin{figure}[t]
\begin{center}
\epsfig{file=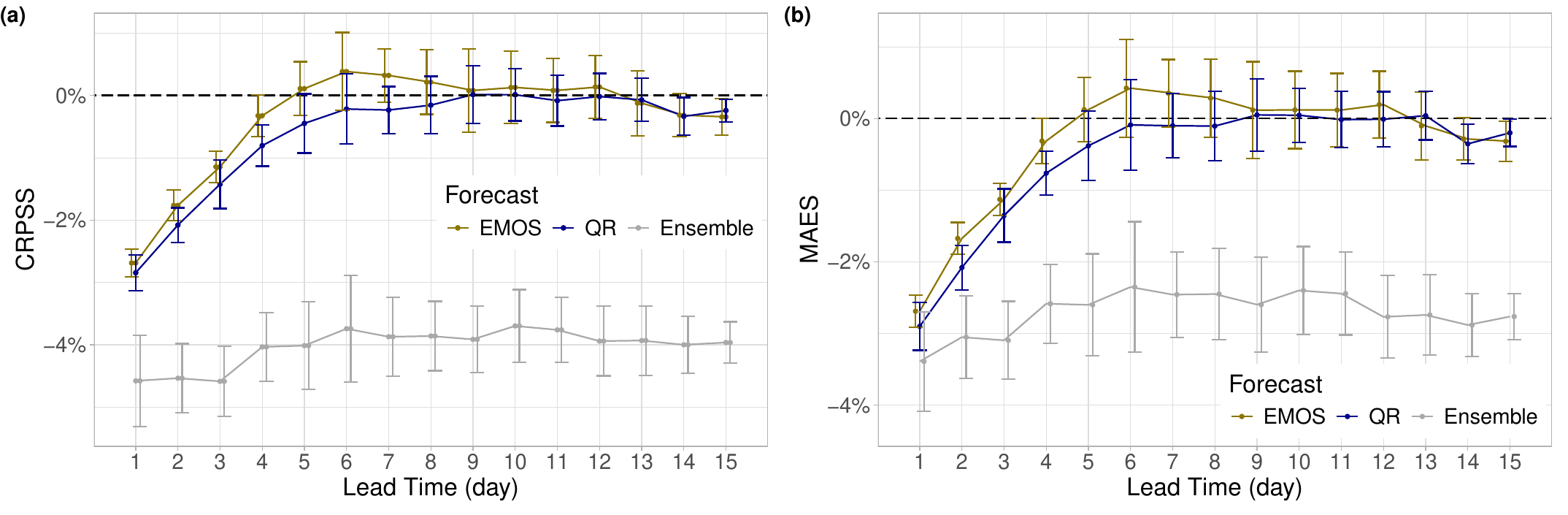, width=\textwidth}
\end{center}
\caption{CRPSS (a) and MAES (b) of post-processed and raw AIFS wind-speed forecasts with respect to the corresponding IFS predictions together with 95\,\% confidence bounds as functions of the lead time.}
\label{fig:crpss_maes_IFS}
\end{figure}

A similar message is conveyed by Figure \ref{fig:mae_maes}; however, the advantage of climatology in terms of the MAE of the median forecast over the raw IFS and AIFS ensemble predictions, for short lead times, is much smaller than in terms of the mean CRPS.

\begin{figure}[t!]
\begin{center}
\epsfig{file=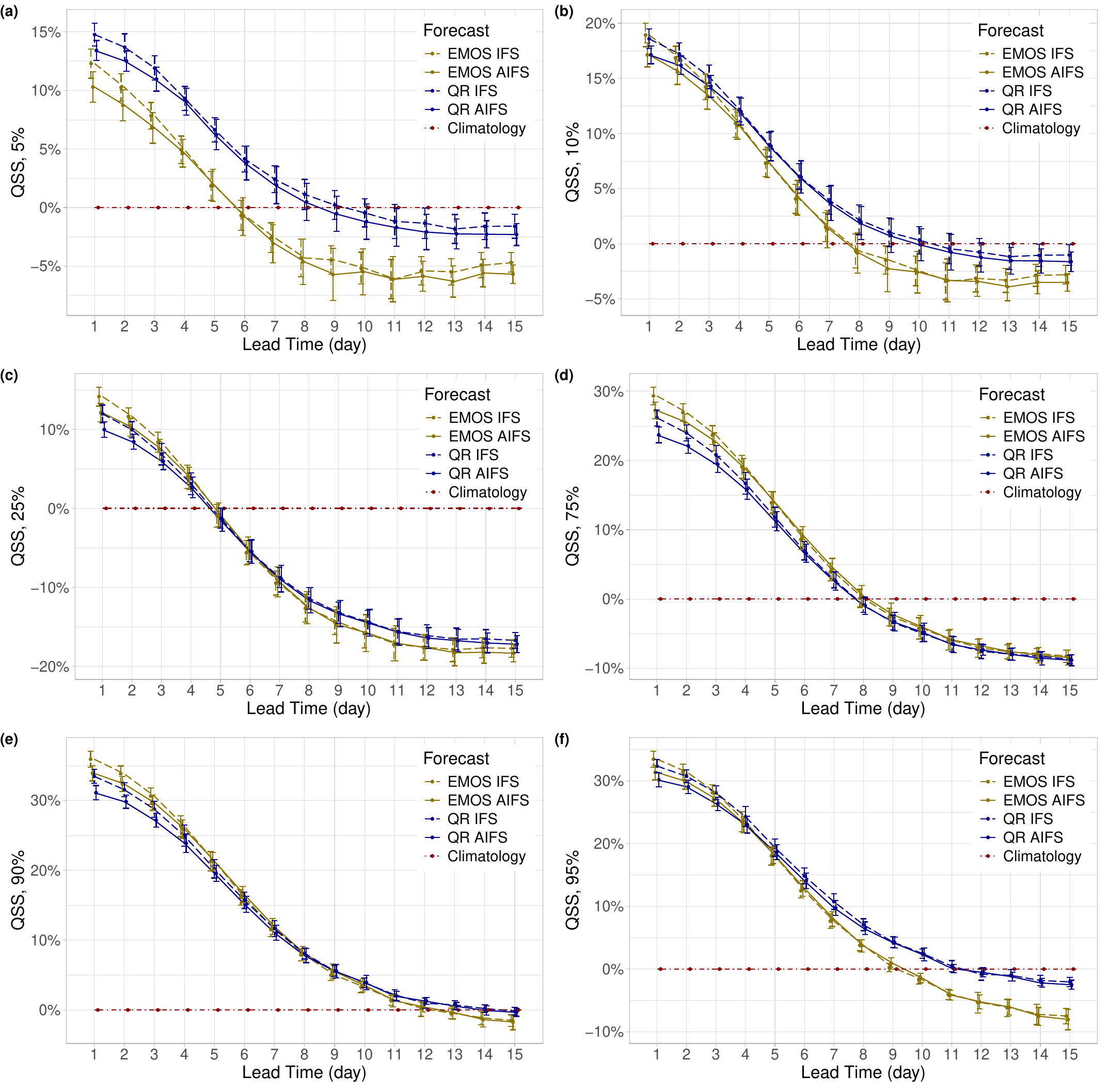, width=\textwidth}
\end{center}
\caption{QSS of post-processed wind-speed forecasts with respect to climatology for percentiles 5 (a), 10 (b), 25 (c), 75 (d), 90 (e) and 95 (f), together with 95\,\% confidence bounds as functions of the lead time.}
\label{fig:qss}
\end{figure}

Figure \ref{fig:crpss_maes_IFS} highlights the differences between the IFS and AIFS forecasts, for both raw and post-processed predictions. In the reported skill scores, the reference is always the matching IFS forecast, e.g., EMOS IFS for EMOS and QR IFS for quantile regression. Thus, Figure \ref{fig:crpss_maes_IFS} does not provide information about the ranking of the three different forecasts.
The raw AIFS forecast significantly underperforms the raw IFS prediction both in terms of the mean CRPS and the MAE, uniformly for all lead times (see also Figures \ref{fig:crps_crpss}a and \ref{fig:mae_maes}a), with CRPSS around -4\,\% and MAES around -3\,\%. Both post-processing techniques drastically change the situation. For both scores, the advantage of the EMOS IFS over the EMOS AIFS is significant up to day 4, showing a decreasing trend with the increase of the lead time, and at day 15, whereas in the middle (days 5 -- 12), the CRPSS and MAES values of the AIFS EMOS are slightly, but not significantly positive. The same behaviour can be observed in the relation of the QR IFS and QR AIFS, although the advantage of the former, both in terms of the mean CRPS and MAE, is a bit greater than in the case of the EMOS.

Note that a location-specific assessment of the significance of differences in mean CRPS between the corresponding IFS and AIFS forecasts can be found in Appendix \ref{secA}.

\begin{figure}[t!]
\begin{center}
\epsfig{file=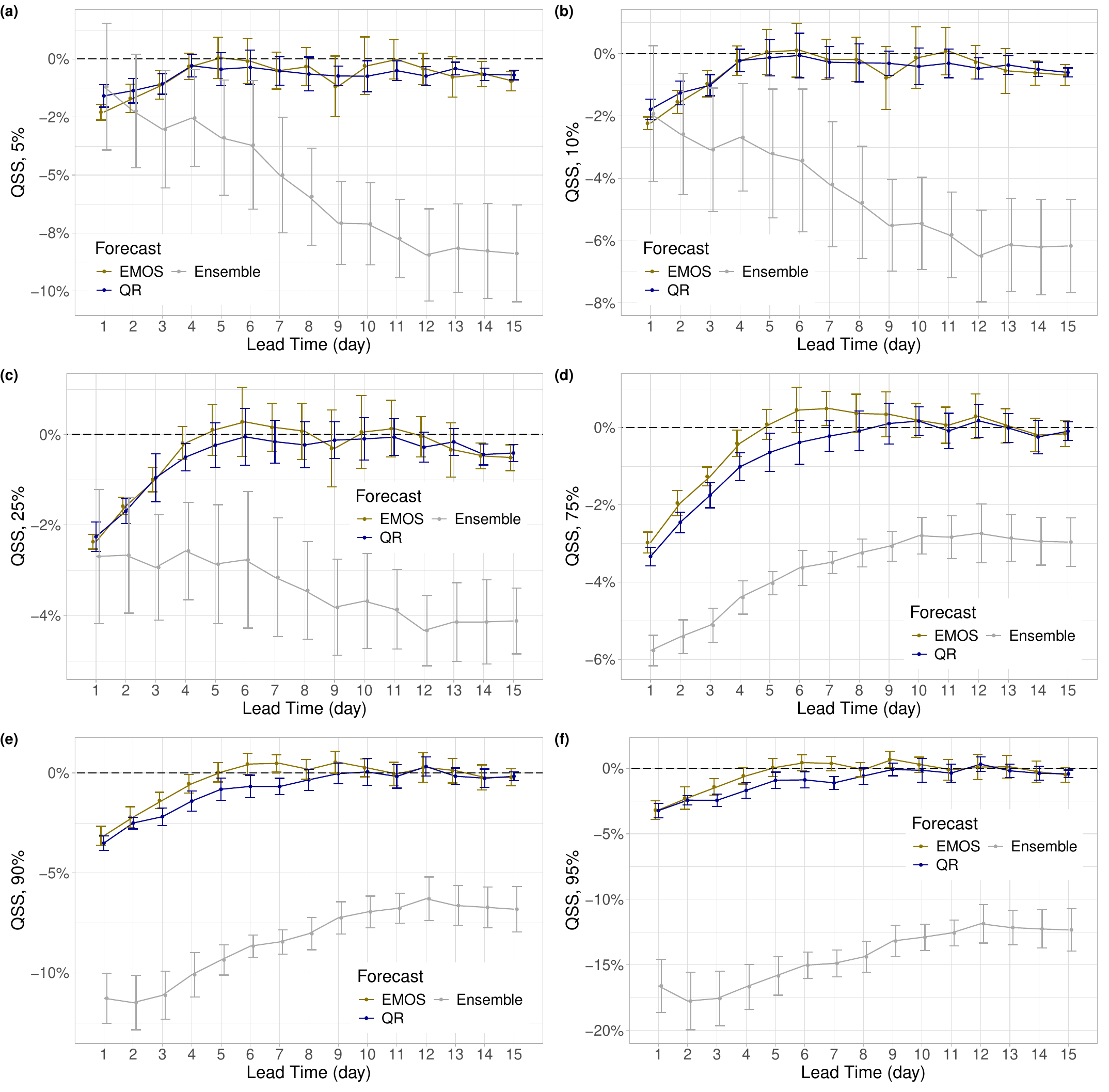, width=\textwidth}
\end{center}
\caption{QSS of post-processed and raw AIFS wind-speed forecasts with respect to the corresponding IFS predictions for percentiles 5 (a), 10 (b), 25 (c), 75 (d), 90 (e) and 95 (f), together with 95\,\% confidence bounds as functions of the lead time.}
\label{fig:qss_IFS}
\end{figure}

Figure \ref{fig:qss} displays the QSS of all post-processed forecasts with respect to climatology. Note that the mean QS values of raw IFS and AIFS forecasts, similar to their mean CRPS, are substantially higher than those of the post-processed predictions for all forecast horizons; hence, they are excluded from the analysis. In general, the skill of post-processed forecasts decreases with the increase of the lead time, while the ranking of the two post-processing methods strongly depends on the considered percentile. At 5\,\% (Figure \ref{fig:qss}a), QR is significantly ahead of the EMOS for all lead times, both for IFS and AIFS predictions; the latter outperforms climatology significantly only till day 5. At 10\,\% (Figure \ref{fig:qss}b), the general advantage of QR still remains; all skill scores are slightly higher, and the period with significantly positive QSS for all four post-processed forecasts is longer. For the other four studied percentiles, at short lead times, EMOS tends to outperform QR, but except 75\,\%, the ranking changes for longer forecast horizons. For the highest two percentiles of 90\,\% and 95\,\%, the QSS at day 1 for all calibrated predictions is above 30\,\% and remains significantly positive untill day 11 and day 8, respectively. Finally, at the 95th percentile, after day 7, EMOS significantly lags behind QR, and the difference increases considerably with the lead time.

\begin{figure}[t!]
\begin{center}
\epsfig{file=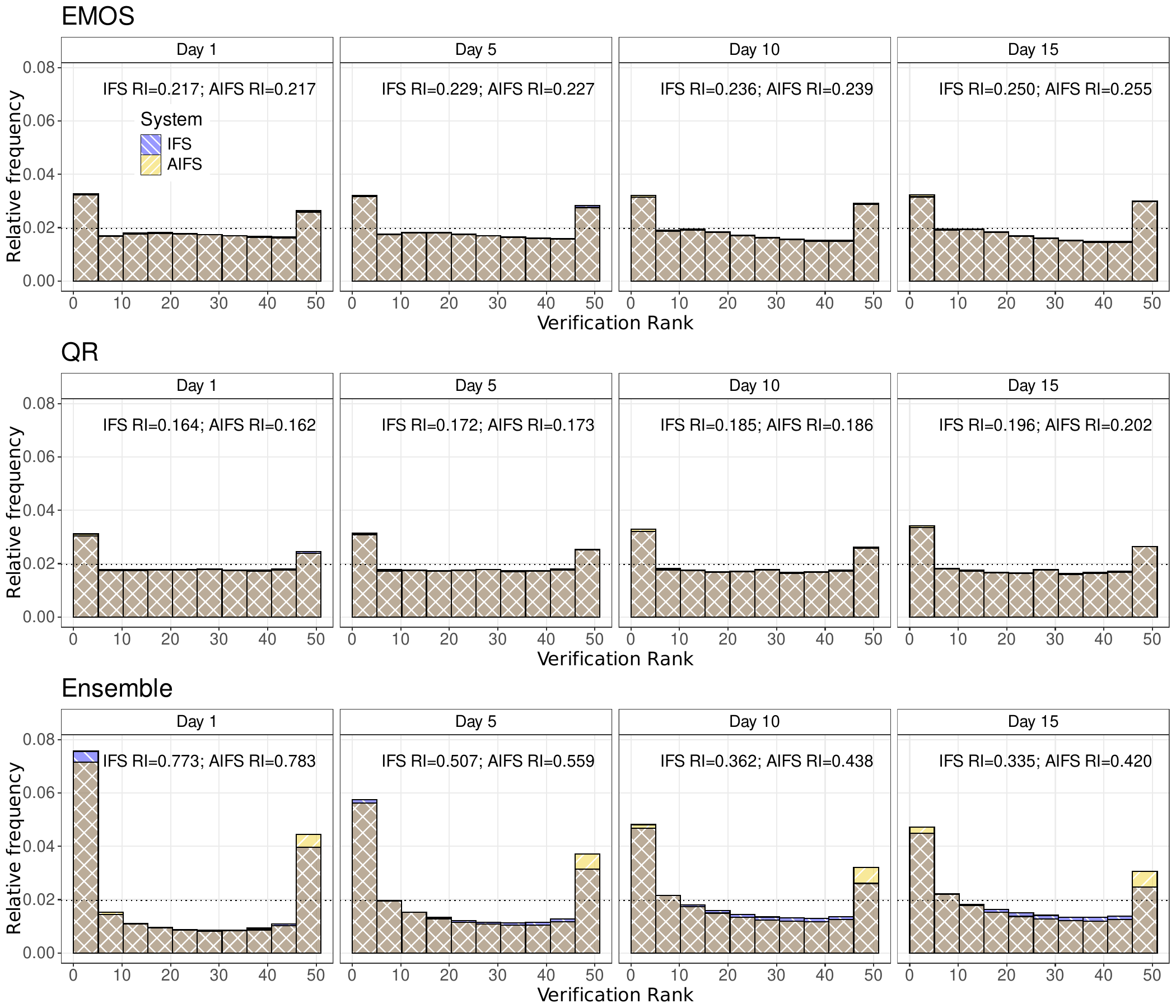, width=\textwidth}
\end{center}
\caption{Verification rank histograms of post-processed and raw wind-speed forecasts for lead times 1, 5, 10, and 15 days. The ideal uniform distribution is indicated by the horizontal dotted line. The cross-hatched parts indicate the intersection of the corresponding histograms for IFS and AIFS predictions.}
\label{fig:pit}
\end{figure}

Following the same structure as Figure \ref{fig:crpss_maes_IFS}, Figure \ref{fig:qss_IFS} focuses on the differences between the physics-based forecasts and their ML-based counterparts. The QSS values of the raw AIFS predictions with respect to the raw IFS forecasts are negative for all lead times and all studied percentiles, although their behaviour differs between low (5\,\%, 10\,\%, 25\,\%) and high (75\,\%, 90\,\%, 95\,\%) percentiles. For the former, the difference in QS between raw AIFS and IFS increases with the lead time, with being non-significant at the first two days at 5\,\% and at day 1 at 10\,\%, while for the latter, QSS values of raw forecasts show an increasing trend. Similar to CRPS and MAE, both post-processing methods reduce the difference in QS between the two forecast types. At low percentiles, the EMOS IFS is significantly ahead of the EMOS AIFS at the first three and the last two lead times, otherwise; while the skill score is sometimes slightly positive, the difference is not significant. The QSS values of QR AIFS with repect to QR IFS show the same pattern; nevertheless, they are negative for all forecast horizons. At high percentiles, the advantage of the IFS at the longest two lead times, in general, is not significant anymore, both for EMOS and QR (except QR at day 15 for the 95th percentile); however, for the latter, the higher the considered percentile, the larger the forecast horizon until the QSS values are significantly negative (day 5 at 75\,\%, day 7 at 90\,\%, day 6 at 95\,\%). 

\begin{figure}[t]
\begin{center}
\epsfig{file=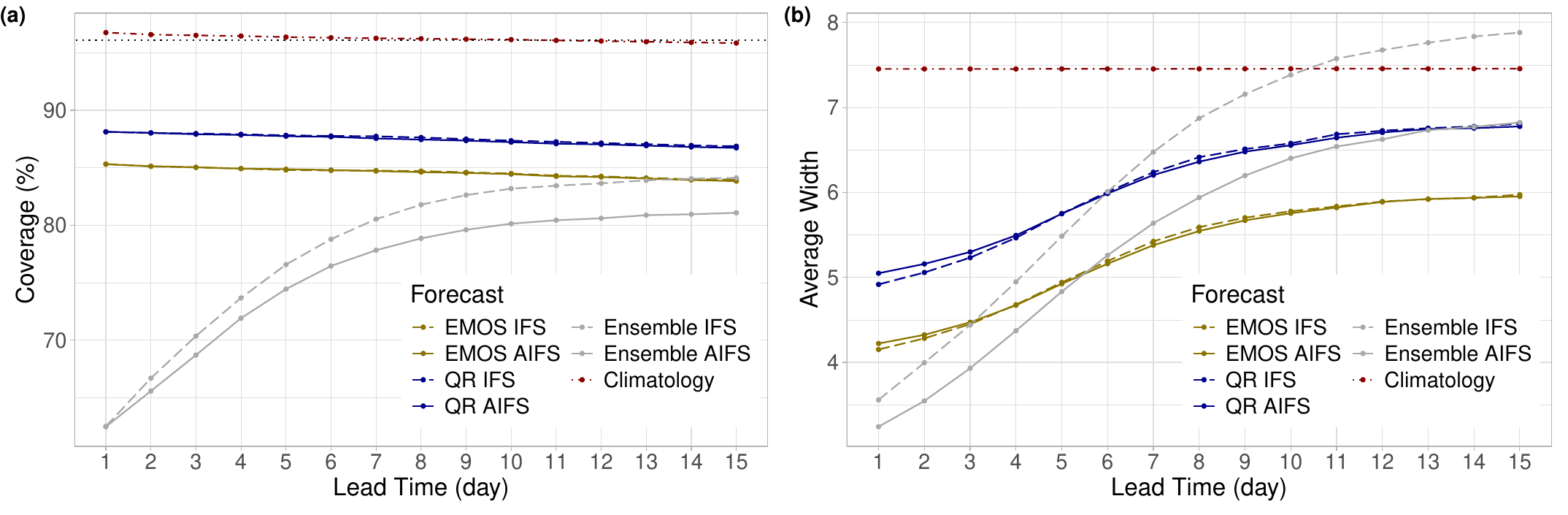, width=\textwidth}
\end{center}
\caption{Coverage (a) and average width (b) of nominal 96.08\,\% central prediction intervals of post-processed, raw and climatological wind-speed  forecasts as functions of the lead time.}
\label{fig:cov_aw}
\end{figure}

Figure \ref{fig:pit} displays the verification rank histograms of raw and post-processed IFS and AIFS forecasts for days 1, 5, 10, and 15, and the corresponding reliability indices. Both the physics-based and the ML-based raw ensemble predictions are strongly underdispersive, resulting in U-shaped rank histograms; moreover, they are slightly biased towards the lower bins. Both deficiencies are reduced with the increase of the forecast horizon, which is also quantified in the decrease of the reliability indices, with the IFS ranks being systematically closer to the uniform distribution than the AIFS ones. Both post-processing methods substantially reduce the underdispersion and the bias, but cannot completely eliminate them, which aligns with the findings of \citet{bszsz21} and  \citet{bl24}, where the TN EMOS approach was considered. The differences between the corresponding histograms of post-processed IFS and AIFS are hardly visible, and this small deviation is further confirmed by the almost identical reliability indices. Finally, from the two competing methods, QR yields flatter rank histograms and lower RI values for all studied forecast horizons.

The improved calibration of post-processed forecasts and the general advantage of QR over EMOS can also be clearly observed on the coverage values of Figure \ref{fig:cov_aw}a. As expected, the coverage of climatology is almost perfect, and due to the increase in the uncertainty of raw forecasts with lead time resulting in a larger spread, both IFS and AIFS coverage values show an increasing trend, with the former being consistently higher. This behaviour is fully in line with the shapes of the matching rank histograms and the corresponding RI values. In contrast, coverage values of the post-processed predictions show slightly decreasing trends and no visible differences between the matching IFS and AIFS predictions, which is again consistent with the related reliability indices in Figure \ref{fig:pit}.

Figure \ref{fig:cov_aw}b further confirms that the improved coverage of the raw ensemble forecasts for long lead times is the result of increased ensemble spread; after day 10, the average width of the raw IFS prediction intervals is even larger than that of the reference climatology. There are just small differences in the average widths of the matching IFS and AIFS post-processed predictions, mainly for short lead times, with the IFS being slightly sharper. Furthermore, for longer lead times, the better calibration of post-processed forecasts is accompanied by improved sharpness, thereby fulfilling the main goal of post-processing \citep{gr07}. Note that a similar behaviour can be observed for raw and EMOS post-processed IFS wind speed forecasts investigated in \citet{bl24}.

\section{Discussion and conclusions}
\label{sec5}

We provided a detailed comparison of the forecast skill of ECMWF's physics-based IFS and AI-based AIFS-CRPS 50-member medium-range 10-m wind-speed ensemble forecasts, focusing both on raw and post-processed predictions. For post-processing, we utilized the local version of the parametric EMOS relying on a truncated normal distribution, and the clustering-based semi-local version of spline-based quantile regression with adjustments to avoid quantile crossing. All forecasts were evaluated against SYNOP observations.

In general, the predictive performance of raw IFS ensemble forecasts proves to be substantially superior to the skill of the raw AIFS predictions for all investigated forecast horizons, which is quantified in significantly lower mean CRPS and mean QS values and better coverage of probabilistic forecasts and lower MAE of the ensemble median. This general advantage of the IFS might partially be explained by the difference in spatial resolution of the two EPSs (9 km vs. 30 km). A natural question to ask, is how this result relates to the corresponding findings of \citet{aifs-crps26}. According to the reported medium-range scorecards for AIFS-CRPS relative to the operational IFS
ensemble \citep[][Figure 4b]{aifs-crps26}, for all forecast horizons, in terms of the mean CRPS, AIFS significantly outperformed IFS at a 5\,\% level in the northern extra-tropics (20$^\circ$N -- 90$^\circ$N) and tropics (20$^\circ$S -- 20$^\circ$N), but underperformed in the southern extra-tropics (20$^\circ$S -- 90$^\circ$S). In contrast, our location-wise analysis (Table \ref{tab1} in Appendix \ref{secA}) reveals, that in all of these three regions and for all lead times, there are substantially more SYNOP stations, where raw IFS outperforms the raw AIFS than the other way around. However, in \citet{aifs-crps26}, forecasts for an earlier but longer time interval were considered and issued for a different, possibly more evenly distributed across regions, set of SYNOP stations.

Post-processing drastically changes the picture. On the one hand, both investigated methods significantly improve the calibration of probabilistic and accuracy of point forecasts, and, across most considered verification metrics, EMOS is superior to QR, especially for short lead times. This could be explained by the different training schemes: the low number of parameters in the EMOS model allows local training, whereas for QR, clustering-based semi-local training seems to be the best choice. On the other hand, the differences in skill between the matching IFS and AIFS predictions (EMOS IFS vs. EMOS AIFS and QR IFS vs. QR AIFS) are substantially decreased and significant only at short lead times and in some cases at days 14 and 15. In all of these cases, the IFS forecast outperforms its AIFS counterpart.

Finally, the additional results presented in Appendix \ref{secA} do not indicate any specific spatial pattern in which either of the two forecast types (IFS or AIFS), in either raw or post-processed form, is consistently superior to its matching pair.

The present study is just a first step towards a comprehensive assessment of the skill of ECMWF's AI-generated ensemble forecasts, leaving open several avenues for further research. First of all, our results rely on data for a relatively short time period, not involving all seasons. For this resaon, here we consider just the simplest parametric and non-parametric post-processing methods not requiring large training datasets, and do not apply advanced ML-based approaches relying also on forecasts of other related weather quantities and location- and time-specific information and requiring large training datasets, such as DRN \citep{rl18} or non-crossing quantile regression neural networks \citep{sy24ncqrnn}. Furthermore, the increasing importance of renewable energy production makes it essential to study the performance of AI-based forecasts of related quantities like 100-m wind speed or solar irradiance, together with the generated probabilistic power forecasts \citep{mblhyb26}. One can also investigate the skill of raw and post-processed AIFS-based probabilistic forecasts of derived user-oriented variables, such as various heat indices \citep{bbpbb20} or wind damage. Finally, the recent model upgrades of 12 May 2026 of the IFS from Cycle49r1 to Cycle50r1\footnote{\url{https://confluence.ecmwf.int/display/FCST/Implementation+of+IFS+Cycle+50r1}} and AIFS ENS from v1 to v2\footnote{\url{https://confluence.ecmwf.int/display/FCST/Implementation+of+AIFS+ENS+v2}} also motivate new comparative studies as soon as a decent amount of forecast data is available.

\bigskip

\noindent
{\bf Acknowledgements} \ This work was supported by the EK\"OP-25-3-I University Research Scholarship Program of the Ministry for Culture and Innovation, funded by the National Research, Development and Innovation Fund. S\'andor Baran acknowledges the support of the National Research, Development, and Innovation Office under grant no. K142849.  Furthermore, the authors are indebted to Martin Leutbecher for providing the ECMWF wind-speed data.

\bibliographystyle{rss}

\bibliography{AIFS_paper}

\begin{appendix}
  \section{Spatial patterns in forecast skill}
  \label{secA}

\begin{figure}[ht!]
\begin{center}
\epsfig{file=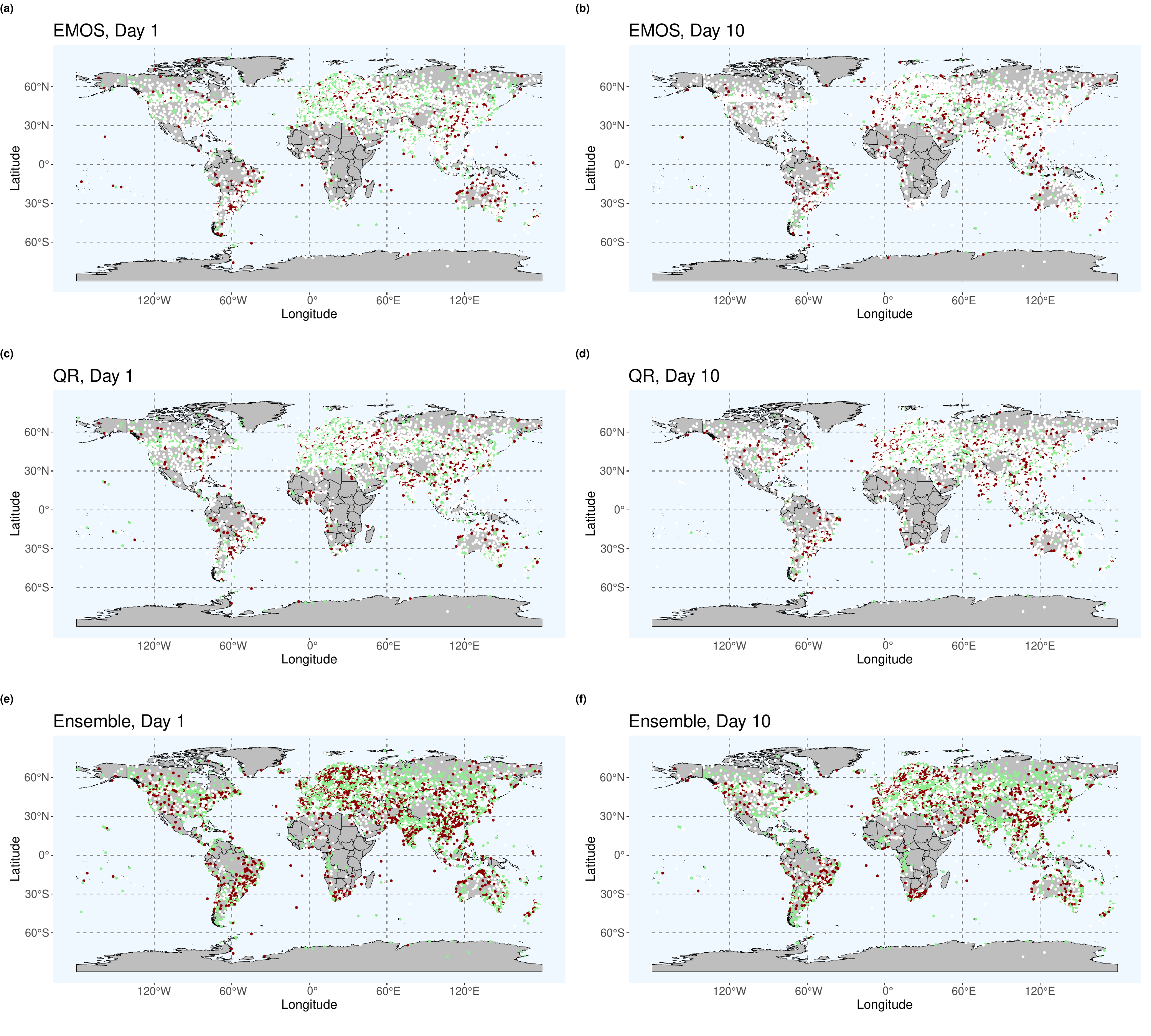, width=\textwidth}
\end{center}
\caption{Location-specific assessment of significance of differences in mean CRPS at a 5\,\% level between (a,b) EMOS and (c,d) QR post-processed and (e,f) raw IFS and AIFS forecasts at (a,c,e) day 1 and (b,d,f) day 5. White dot: the difference is non-significant; light (green) dot: IFS significantly outperforms AIFS; dark (red) dot: AIFS significantly outperforms IFS.}
\label{fig:spatialMap}
\end{figure}

To reveal possible spatial patterns in the relative performance of matching IFS and AIFS forecasts in terms of the mean CRPS, for each location and lead time, we perform pairwise Diebold-Mariano \citep[DM;][]{dm95} tests of equal predictive performance. Figure \ref{fig:spatialMap} displays the results of the one-sided DM tests at a 5\,\% level of significance for all three pairs of forecasts at days 1 and 10 in the form of color-coded maps, where SYNOP stations with no significant difference between IFS and AIFS forecasts are white dots, locations where IFS significantly outperforms AIFS are light (green) dots, and places with AIFS resulting in significantly lower mean CRPS than IFS are dark (red) dots.

\begin{table}[t]
  {\scriptsize
    \begin{tabular}{lccccccccccccccc} \hline
      \multicolumn{16}{c}{Northern Extra-Tropics (20$^\circ$N -- 90$^\circ$N), 7533 SYNOP Stations}\\\hline
      Superior&\multicolumn{15}{c}{Lead Time (day)}\\ \cline{2-16}
      Forecast &1&2&3&4&5&6&7&8&9&10&11&12&13&14&15 \\ \hline
   EMOS IFS&19.4&15.6&13.6&10.8&8.7&7.5&8.3&7.8&7.8&8.4&8.1&7.1&7.3&8.1&8.1 \\
      EMOS AIFS&6.3&7.0&7.3&9.1&9.4&9.7&8.9&8.2&8.3&7.2&6.7&7.3&6.9&6.0&5.7\\ \hline
    QR IFS&18.9&16.6&14.7&12.6&10.9&9.9&10.5&10.1&9.7&10.1&10.2&9.7&9.7&10.2&10.5\\
       QR AIFS&7.2&7.8&8.1&8.8&8.8&9.3&8.9&8.3&9.3&8.9&9.6&9.0&8.9&7.7&8.2\\ \hline
      Ensemble IFS&34.6&32.6&31.8&29.5&28.5&26.9&27.8&28.2&28.6&28.4&28.5&28.9&28.9&29.3&30.2\\
      Ensemble AIFS&28.8&27.1&26.2&26.0&24.7&24.2&23.0&22.1&22.8&23.2&23.7&21.6&21.7&21.1&20.7 \\ \hline \hline
\multicolumn{16}{c}{Southern Extra-Tropics (20$^\circ$S -- 90$^\circ$S), 914 SYNOP Stations}\\\hline
      Superior&\multicolumn{15}{c}{Lead Time (day)}\\ \cline{2-16}
      Forecast (\%) &1&2&3&4&5&6&7&8&9&10&11&12&13&14&15 \\ \hline
   EMOS IFS&13.2&12.4&11.7&12.0&10.2&10.1&10.2&9.3&7.0&7.9&7.4&5.7&6.0&5.9&6.9 \\
      EMOS AIFS&9.8&12.4&13.2&11.2&11.7&9.6&10.8&10.9&8.6&8.6&8.3&10.3&8.9&7.1&10.2\\ \hline
    QR IFS&17.1&13.7&13.2&13.9&11.9&14.7&11.3&11.7&10.5&8.9&8.0&8.0&8.1&8.4&7.9\\
       QR AIFS&11.6&10.9&12.8&11.5&10.9&10.4&9.7&11.1&11.1&11.9&9.5&10.5&10.3&8.4&9.8\\ \hline
      Ensemble IFS&36.7&33.7&32.4&34.2&34.0&34.2&33.6&33.5&32.6&31.0&31.4&31.5& 30.5&31.9&30.2\\
      Ensemble AIFS&25.9&24.6&25.2&24.6&24.0&22.9&24.8&26.4&25.8&25.1&24.5&26.7&25.5&24.2&23.3 \\ \hline \hline
\multicolumn{16}{c}{Tropics (20$^\circ$S -- 20$^\circ$N), 799 SYNOP Stations}\\\hline
      Superior&\multicolumn{15}{c}{Lead Time (day)}\\ \cline{2-16}
      Forecast (\%) &1&2&3&4&5&6&7&8&9&10&11&12&13&14&15 \\ \hline
   EMOS IFS&8.5&5.9&5.5&5.6&5.5&4.5&3.9&3.5&3.8&5.0&3.8&4.4&4.8&3.9&5.9 \\
      EMOS AIFS&11.6&10.3&12.0&12.1&11.4&13.4&12.6&12.9&14.1&14.0&12.9&11.5&9.9&8.9&9.6\\ \hline
    QR IFS&11.0&9.6&6.6&9.6&7.1&7.3&5.8&7.0&5.6&7.1&8.1&6.9&8.0&9.6&8.1\\
       QR AIFS&13.3&12.5&12.9&13.1&13.4&14.9&13.4&15.4&11.8&11.6&12.9&11.9&11.8&11.6&11.3\\ \hline
      Ensemble IFS&36.2&34.4&34.8&34.7&34.8&33.8&34.3&33.9&33.7&34.0&33.5&34.9&35.3&35.0&33.0\\
      Ensemble AIFS&36.0&31.2&30.7&28.7&27.0&26.8&25.4&25.8&24.5&23.5&24.7&23.8&24.8&24.9&25.4 \\ \hline
    \end{tabular}}
  \caption{Proportion of SYNOP stations, where the IFS and the AIFS are significantly superior in terms of the mean CRPS.}
  \label{tab1}
\end{table}

None of the presented maps shows a clearly identified spatial pattern when one of the forecast types (IFS or AIFS) is superior to the other. Nevertheless, they nicely align with the behaviour of the CRPSS in Figure \ref{fig:crpss_maes_IFS}a. At day 1, the difference between EMOS IFS and EMOS AIFS forecast is significant for 25.0\,\% of SYNOP stations, with the IFS being significantly superior in 71\,\% of the cases. The proportion of locations with significant differences in mean CRPS reduces to 16.0\,\% by day 10, and the share of SYNOP stations with significant IFS lead decreases to 50.4\,\%. Note that the corresponding overall CRPSS values are -2.687\,\% and 0.128\/\%, respectively. For the QR, the share of SYNOP stations where the difference between the two types of forecasts is significant is 26.2\,\% at day 1 (IFS lead: 68.9\,\%) and 19.2\,\% at day 10 (IFS lead: 50.8\,\%); the corresponding overall skill scores are -2.844\,\% and 0.011\,\%, respectively. The overall CRPSS values of the raw AIFS forecasts with respect to the IFS ensemble are -4.578\,\% at day 1 and -3.687\,\% at day 10, suggesting more locations with significantly different mean CRPS. Indeed, at day 1, this proportion is 64.1\,\% (IFS lead: 54.5\,\%), while at day 10, it is 52.6\,\% (IFS lead: 55.5\,\%).

A more detailed picture can be obtained from Table \ref{tab1}, where, following \citet{aifs-crps26}, the northern and southern extra-tropics (20$^\circ$N -- 90$^\circ$N and 20$^\circ$S -- 90$^\circ$S), and the tropics (20$^\circ$S -- 20$^\circ$N) are considered separately. For each region, it reports for each forecast separately the proportion of SYNOP stations among all considered locations, where the IFS and the AIFS are significantly superior in terms of the mean CRPS. Raw IFS is ahead of the raw AIFS in all three regions and for all lead times. Post-processing substantially changes this clear ranking. At the tropics, both EMOS AIFS and QR AIFS are ahead of their IFS counterparts for each forecast horizon. QR IFS clearly defeats QR AIFS at the northern extra-tropics and up to day 9 at the southern extra-tropics. Finally, EMOS IFS wins at more locations in the northern extra-tropics than EMOS AIFS up to day 4 and after day 10, whereas in the southern extra-tropics it clearly loses its advantage  after day 6. However, one should also note that the distribution of SYNOP stations between the three regions is highly unbalanced, and for the post-processed forecasts, the difference in mean CRPS between the matching IFS and AIFS forecasts fails to be significant for more than 70\,\% of the locations.
\end{appendix}

\end{document}